\documentclass[aps,prb,twocolumn,groupedaddress]{revtex4}

\usepackage{graphicx}
\usepackage{bm}

\begin{document}


\title{Charge fluctuations in cuprate superconductors}
\author{Zhongqing Wu$^1$ and Xuanjia Zhang$^2$}
\affiliation{$^1$Department of Physics,Tsinghua University, Beijing 100084,China\\
$^2$Department of Physics, Zhejiang University, Hangzhou 310027, China}
\date{\today}

\begin{abstract}
 The effect of the lattice periodic potential on
superconductivity which was ignored by BCS theory has been
investigated. According to the effective mass approximation of
band theory, the effect of lattice periodic potential can be
embodied in the effective mass $m^{\ast }$ of the Ginzburg-Landau
(GL) equations. A special property of $m^{\ast }$ is that it can
be negative. Negative effective mass leads to many unusual
phenomena. The superconducting order parameter shows the period
distribution. Its modulate wavevector is proportional to the
condensed carrier density, which explains the linear relation
between the magnetic peaks displacement $\epsilon $ and $x$ for
La$_{2-x}$Sr$_x$CuO$_4$. The superconducting phase is always local
and separated originally and evolves into global superconducting
phase at the certain pairs concentration, which explains why the
cuprate superconductors must be insulator at low doped. The doped
concentrations of insulator to superconductor
transition for La$_{2-x}$Sr$_x$CuO$%
_4 $ is consistent with the experiment results.  The relation of
the superconducting gap (SG) and the pseudogap (PG) was discussed.
\\
Keywords: Stripes; Ginzburg-Landau theory; Superconductivity;
Charge fluctuations.
\end{abstract}

\pacs{PACS: 74.80.-g, 74.20.De, 74.25.Dw }


\maketitle


Since the high temperature superconductor (HTS) was found in 1986,
HTS has been one of the most attractive field in condensed matter
physics. Almost any physical properties about HTS have been
intensely studied and many anomalous properties have been found.
which make many people believe that the BCS theory and Fermi
liquid theory are no longer fit for HTS. The most striking of all
anomalous properties are the stripe phase and pseudogap (PG).
Unlike conventional metals in which the charge distribution is
homogenous, the charge carriers are segregated into
one-dimensional stripe in HTS, which was predicted by t-j
model\cite{Zan1} and then supported by many
experiment\cite{Moo,Dai,Nod,Yam,Mat,Tran,Tranq}. The relations of
many physical properties such as Knight Shifts\cite{Wal}, NMR
relaxation rates\cite{Ish}, DC conductivity\cite{Mom},
 with the temperature in normal state of HTS do not conform to Fermi liquid theory,
which has been valid in the conventional metal. The temperature at
which these physical properties begin to depart from Fermi liquid
theory are almost the same. Some experiments such as
angle-resolved photoemission, tunnelling spectroscopy indicate
that the density of states (DOS) in Fermi surface begin to
decrease and an PG open at this temperature. The PG is of the same
size and k dependence as superconducting gap (SG). The microscope
mechanism of PG is still unclear. In superconducting states of
HTS, anomalous properties have also been found. For example, the
ratio of the energy gap in 0K $\Delta(0)$ with the superconducting
phase transition temperature $T_c$ ($2\Delta(0)/k _{B}T
_{c}\approx 8\sim9$) far more than BCS theory' prediction
($2\Delta(0)/k _{B}T _{c}=3.53$)

Some people believe that these anomalous properties result from
the correlation effect between electrons, which has been overlook
by BCS and Fermi liquid theory. Many model have been proposed for
the study of correlation effects. However up to now, there is a
lack of the breakthrough in this aspect.

Another important effect that has also been overlook by BCS theory
is the lattice periodic potential. The band theory that describes
the effect forms the basis of the modern theory of electron in
solids. According to the effective mass approximation of band
theory, the effect can be included in BCS theory if we substitute
the effective mass $m^{\ast }$ for mass of bare electron. $m^{\ast
}$ should be negative for HTS, as will be discussed in the end of
our paper. Since Ginzburg-Landau (GL) equations can be derived
from the BCS theory, the mass in equation should be also negative
for HTS. Negative effective mass will lead to extraordinary
spatial distribution of the order parameter, which can well
illuminate many anomalous properties.

Localized holes organize into one-dimensional structures, which has been observed in HTS in many experiments\cite%
{Tran,Bor,Nod,Tranq,Sha}. The spin modulation also shows
one-dimensional properties. Neutron scattering studies reveal that
there are two types of
twin domains and spin modulation is one dimensional in each domain for La$%
_{2-x}$Sr$_{x}$CuO$_{4}$ and YBa$_2$Cu$_3$O$_6.6$\cite{Tra,Moo}.
We think that these one-dimensional properities result from the
one-directional modulation of order parameter. Then with no
magnetic field considered, the first equation of GL can be written
as
$$
\alpha \psi +\beta {|\psi |}^{2}\psi -\frac{\hbar ^{2}}{2m_{l}^{\ast }}%
\frac{\partial^{2} \psi}{\partial l^{2}} =0\text{,}\eqno{(1)}
$$%
where $m_{l}^{\ast }$ is the component of the effective mass in
$l$ direction, $\psi =\sqrt{n_{s}(r)}e^{i\phi (r)}$ is the order
parameter, which only vary along $l$ and is equal in direction
perpendicular to $l$, where $n_{s}(r)$ is the local condensed
carrier density and $\phi $ is the phase of the effective wave
function.

Taking $f=\frac{\psi }{\psi _{0}}$ as the dimensionless effective wave
function where $|\psi _{0}|^{2}=-\frac{\alpha }{\beta }$,\ under the
condition $m_{l}^{\ast }<0$, the equation (1) reduces to
$$
\xi ^{2}\frac{\partial^{2}f}{\partial
l{}^{2}}=f(1-f^{2})\text{,}\eqno{(2)}
$$%
where$\ \xi ^{2}=\frac{\hbar ^{2}}{|{2m_{l}^{\ast }\alpha }|}$.
The corresponding equation for $m_{l}^{\ast }>0$ can be obtain as
long as the left side of equation (2) multiply by -1. As will be
shown later, just above difference between equation for
$m_{l}^{\ast }>0$ and $m_{l}^{\ast }<0$ leads to that they have
very different spatial distribution of the order parameter. Since
the coefficient of each term in equation (2) does not include the
variable $l$, the origin can be arbitrary for the infinity system.
Therefore we choose the origin where $f$ has the minimum. Then the
boundary condition is
$$
f(0)=f_{0}\text{,}
$$
$$
\left.\frac{\partial f}{\partial l}\right|
_{l=0}=0\text{.}\eqno{(3)}
$$

\begin{figure}[]
\includegraphics[width=8.5cm]{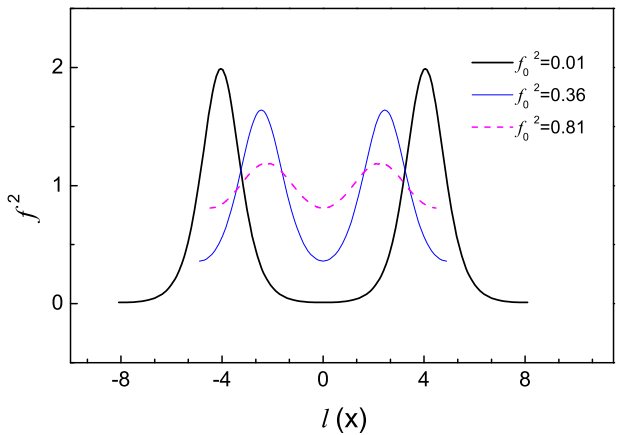}
\caption{$f^{2}=\frac{|\protect\psi |^{2}}{|\protect\psi
_{0}|^{2}}$ as a function of space coordinate $l$ (in units of
$\xi$) for $f_{0}^{2}=0.01,0.36,0.81$} \label{fig:1}
\end{figure}

The dependence of $f^{2}$ on the space coordinate $l$ is shown in
Fig. 1. The spatial distribution of the order parameter is
sensitive to $f_{0}^{2}$. When $f_{0}^{2}$ is close to 1, the
condensed carrier density fluctuation is similar to the charge
density wave. When $f_{0}^{2}$ is close to 0, the phase separation
is obvious. The condensed carriers get together and form
periodically charged stripes. Between the charged stripes are
regions where few condensed carrier can be found and the
microscope composition is similar with the parent compounds of
HTS. We call these regions antiferromagnetic (AF) stripes. It is
shown that AF stripes are wider than charge stripes when
$f_{0}^{2}=0.01$ in Fig. 1, which is more distinct when further
decreasing $f_{0}^{2}$. The stripes have been observed by many
experiments in HTS and their structure change with the doped
concentration. The low-energy neutron-scattering studies
performed on La$_{2-x-y}$Nd$_{y}$Sr$_{x}$CuO$_{4}$ by Yamada$\ {\it {et}}$ $%
{\it {al}}$\cite{Yam} have demonstrated that the charge modulation
wavevector $\epsilon $ initially increases linearly with $x$ before saturating for $%
x>1/8$.

 In order to compare with the results of the low-energy neutron-scattering
studies, we investigated the relation of the average condensed
carrier density with the order parameter modulation wavevector.
The average condensed carrier density is defined as
$$
\rho =\frac{1}{\tau }\int_{-\frac{\tau }{2}}^{\frac{\tau }{2}}{|\psi |}%
^{2}dl=\frac{1}{\tau }{|\psi _{0}|}^{2}\int_{-\frac{\tau }{2}}^{\frac{\tau }{%
2}}|f|^{2}dl,\eqno{(4)}
$$%
where $\tau $ is the order parameter modulation period.
The dependence of $\rho $ on $\frac{1}{\tau }$ is shown in Fig. 2. $\rho $
varies linearly with $\frac{1}{\tau }$ as
$$
\rho =\frac{\xi |\psi _{0}|^{2}}{k}\frac{1}{\tau }\eqno{(5)}
$$%
for $\rho <0.6{|\psi _{0}|}^{2}$, where $k=\frac{4}{15}$ is the
slope of the broken line. Because that the modulation wavevector
$\epsilon $ is proportional to the inverse of the charge
modulation period and the average condensed carrier density $\rho
$ is proportional to Sr concentration $x$ for
La$_{2-x}$Sr$_{x}$CuO$_{4}$ if all of carrier condensed, the
equation (5) denotes
$$
\epsilon \propto x\eqno{(6)}
$$%
when $\rho <0.6{|\psi _{0}|}^{2}$. Furthermore, the modulation
wavevector $\epsilon $ reaches the maximum when the order
parameter is homogenous distribution. These results accord well
with the neutron scattering measurements on
La$_{2-x-y}$Nd$_{y}$Sr$_{x}$CuO$_{4}$\cite{Yam,Mat}.

\begin{figure}[]
\includegraphics[width=8.5cm]{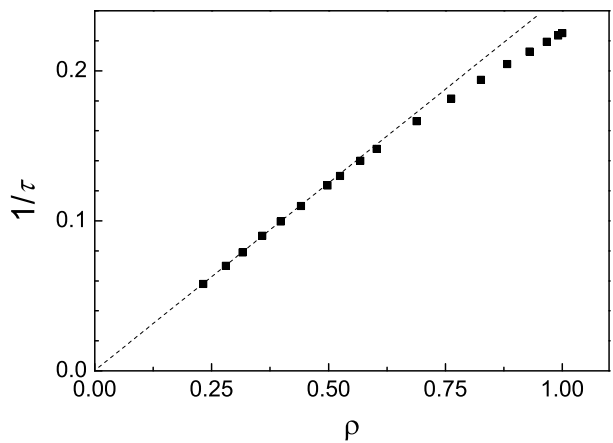}
\caption{The dependence of inverse of charge modulation period $\frac{1}{%
\protect\tau }$(in units of $\frac{1}{\xi}$) on the condensed
carrier density $\protect\rho$ ( in units of $|\psi _{0}|^{2}$).}
\label{fig:2}
\end{figure}

Fig. 1 indicates that there are two kinds of the spatial
distribution of the condensed carriers density: One is the global
distribution, the other shows that the condensed carriers is local
and separate each other. The kind of the distribution can be
determined by whether $f_{0}^{2}$ is close to 0 and should be
close correlative with the average condensed carrier density $\rho
$. The relation of $f_{0}^{2}$ with $\rho $ is shown in Fig. 3.  When $\rho $ is smaller than $%
\rho _{c}=0.35{|\psi _{0}|}^{2}$, $f_{0}^{2}$ is almost 0 and the
superconducting phase is local and separated each other, which
leads to that the system does not show global superconductivity.
It is obvious the temperature of the global superconductivity
$T_c$ is determined by the value of $f_0^2$ at ground state.
Expanding $T_c$ with the value of $f_{0}^{2}$ at ground state by
Thaler Formula, we have that $T_{c}$ is proportional to
$f_{0}^{2}$ when $f_{0}^{2}<<1$. Because all of carrier condensed
at ground state, so $\rho \propto x$. Therefore, in fact, Fig. 3
reveals the dependence of $T_{c}$ on Sr concentration $x$. $\rho
_{c}$ just corresponds to the Sr concentration $x_{c}$ of
non-superconductor to superconductor transition. According to the
fact that modulated
wave vector $\epsilon $ has maximum at $\rho ={%
|\psi _{0}|}^{2}$ in our result and is saturating at $x=1/8$ by
the experiment \cite{Yam}, we evaluated $\ x_{c}$ to be about
$0.044$ for La$_{2-x}$Sr$_{x}$CuO$_{4}$.

\begin{figure}[]
\begin{center}
\includegraphics[width=8.5cm]{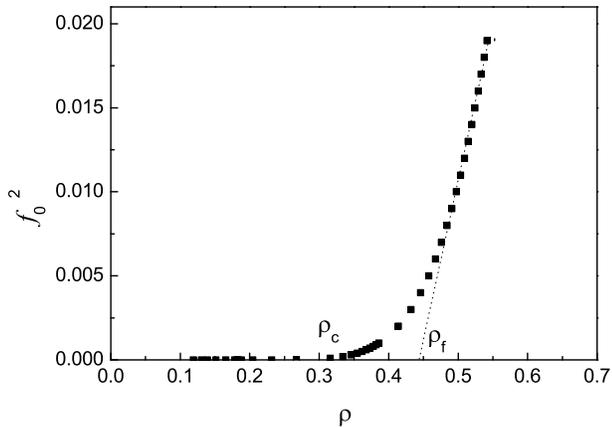}
\end{center}
\caption{$f_{0}^{2}$ as a function of the condensed carrier
density $\protect\rho $ ( in units of $|\psi _{0}|^{2}$)}
\label{fig:3}
\end{figure}

 It is well known that the free energy of system must decrease with increasing the condensed carrier
 density. Obviously, we should make sure whether our free energy meets the above demand.
The free energy density of the system
$$
F_{s}=\frac{1}{\tau }\int_{-\frac{\tau }{2}}^{\frac{\tau
}{2}}[\alpha |\psi |^{2}+\frac{\beta }{2}|\psi
|^{4}+\frac{1}{2m_{l}^{\ast }}|(-i\hbar \nabla )\psi
|^{2}]dl,\eqno{(7)}
$$%
where the free energy density in the normal state is taken as
zero. The relation of $F_{s}$ (in units of $\alpha |\psi
_{0}|^{2}$) and $\rho $ is shown in Fig. 4. $F_{s}$ decrease
monotonously with increasing the condensed carrier density $\rho
$(note: $\alpha $ is negative). It implies that more and more
carriers condensed with decreasing the temperature. An intriguing
phenomenon is that $F_{s}$ is weak-dependence on $\rho $ when
$\rho \approx 1$, which means that the electron pairs can be
excited without energy. Namely, the energy gap has node. At the
conventional superconductor, the energy gap is non-zero and
manifests itself in exponentially activated temperature dependence
of a wide variety of dynamic and thermodynamic properties at low
temperature. At HTS the exponentially activated temperature
dependence is disappear. Cavity perturbation
measurements\cite{Zha} and muon spin rotation study\cite{Har} on
high quality YBa$_2$Cu$_3$O$_7$ crystals have revealed the linear
temperature dependence in the penetration depth $\lambda$ below 30
K. A linear term in the low temperature thermal conductivity has
 also been found\cite{Kri}. Above linear temperature
dependence of physical properties indicates non-exponentially
activated temperature dependence of the DOS of the quasiparticle
excitation, which can appear only when there is node in the energy
gap. The presence of a zero-bias conductance peak in tunneling
spectroscopy also supports that the energy gap has node\cite{Hu}.


\begin{figure}[]
\includegraphics[width=8.5cm]{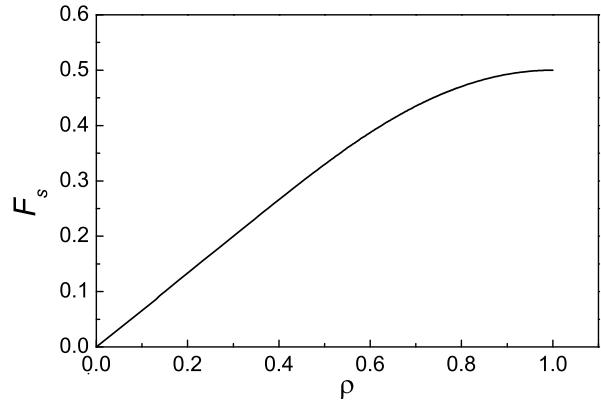}
\caption{The dependence of the free energy density $F_{s}$ (in
units of $\alpha |\psi _{0}|^{2}$ ) on the condensed carrier
density $\protect\rho $ ( in units of $|\psi _{0}|^{2}$).}
\label{fig:4}
\end{figure}

 The electrons begin to pair below the superconducting phase transition
temperature. But the system does not show the zero-resistance
properties because the superconducting phase is local at low pairs
concentration. The condensation of electrons leads to the decrease
of density of states in Fermi surface, which results in the
anomalous of normal state transport properties and the open of the
pseudogap (PG). The normal state PG turns into the superconducting
gap (SG) when the local superconducting phase evolves into the
global one. Therefore it is natural that many experiment, such as
angle-resolved photoemission, tunnelling spectroscopy, nuclear
magnetic resonance and neutron scattering, reveal that SG emerges
from
the normal state PG and is of the same size and k dependence as PG\cite{Tim}%
. Even all of carriers transform into electron pairs, the
superconducting phase is still local and separate for $x<x_{c}$
($=0.044 $ for La$_{2-x}$Sr$_{x}$CuO$_{4}$). The system does not
show global superconductivity at any temperature. They become
insulators because the local superconducting phase consumes all of
carrier at low temperature. For $x>x_{c}$, the local
superconducting phase can always evolve into the global one at
certain temperature. Therefore the $x_{c}$ is the concentration of
insulator to superconductor transition. Since the transition
concentration is determined by extending the experiment data of
$T_{c}\neq 0K$, it may be arbitrary a value between $\rho _{c}$
and $\rho _{f}$. This gives out a likely explanation why the
insulator-superconductor transition concentration reported by
different experiment groups is obvious disagreement. $\rho
_{c}=0.34|\psi _{0}|^{2}$ and $\rho _{f}=0.45|\psi _{0}|^{2}$
respectively correspond to $x=0.044$ and $x=0.056$ for
La$_{2-x}$Sr$_{x}$CuO$_{4}$, which is well consistent with the
experiment result $x\approx 0.05\sim {0.06}$\cite{Tak,ell}.

Contrary with the common sense, the superconducting phase appears
also in insulator. Furthermore the modulate wave vector $\epsilon
$ in insulator and in superconductor has the same dependence on
doped concentration, $\epsilon \propto x$,
which is verified by neutron scattering experiment on La$_{2-x}$Sr$_{x}$CuO$%
_{4}$\cite{Mat}.  The Josephon effect experiment
by Decca ${\it {et}}$ ${\it {%
al}}$\cite{Dec} indicates the existence of an anomalously large
proximity effect in underdoped insulating
YBa$_{2}$Cu$_{3}$O$_{6+x}$. Our results that insulator has local
superconducting phase will be helpful to understand above
anomalously large proximity effect.

 It is generally perceived that the superconducting phase transition
temperature is the temperature $T_C$ at which the resistance
disappear. However, according to our result, this view only suits
the conventional metals. The genuine phase transition temperature
for HTS is the PG temperature $T_p$ where
 the order parameter has appeared. The zero-resistance temperature $T_C$ is just
the temperature at which the local superconducting phase is
translated into the globe superconducting phase. It is well known
that the gap should vanish and the breaking of the symmetry will
happen at the phase transition temperature. However, these
phenomena do not occur at the zero-resistance temperature $T_C$.
On the contrary, a breaking of time-reversal symmetry at the PG
temperature $T_p$ is recently found by angle-resolved
photoemission study\cite{Kam}. Therefore, at the region between
$T_p$ and $T_C$, the system that has been taken for normal state
is actually in superconducting state. Then it is natural that many
physical properties at this region do not conform to Fermi liquid
theory that describes the normal-state metal. Noting that the
physical properties above the PG temperature accord well with
Fermi liquid theory, We are sure that Fermi liquid theory is still
fit for HTS. The opinion that the PG temperature is
superconducting phase transition temperature is helpful to clarify
the perplexity that predict of BCS is invalid for HTS. With the PG
temperature $T_p$ given by the experiments, the value of
$2\Delta(0)/k _{B}T _{p}$ is close to the predict of BCS theory.

Although the negative effective mass $m^{\ast }$ is general
concept in Solid State Physics, it is still necessary to discuss
whether the effective mass may be negative in GL theory. GL theory
is a phenomenological theory. Considering that GL theory can be
derived from the microscopic BCS theory and the meaning of
electron mass in BCS theory is clearer than that in GL theory, we
first study the electron mass of BCS theory. Ashcroft had put
forward that in its simplest form the BCS theory makes a gross
oversimplification in the basic Hamitanian that describes the
conduction electrons \cite{Ash}. The conduction electrons are
treated in the free electron approximation and the effect of
lattice periodic potential (band structure effect) is ignored. The
oversimplifications may seem surprising because the band theory
that describes the effect forms the basis of the modern theory of
electrons in solids. Then the questions why BCS theory that
ignores the effect can still explain the conventional
superconductivity and whether the effect can be neglected in HTS
rise.

 To answer these questions, we must consider
the effect of the lattice periodic potential. The effective mass
approximation of band theory shows that the electrons nearby the
band gap can be treated as free electrons with the effective mass.
The effect of lattice periodic potential is embodied in the
effective mass. According to the approximation, if Fermi surface
is nearby the band gap, the conduction electrons can be taken as
free electron. Then the lattice periodic potential is considered
as long as we replace the mass of bare electron with the effective
mass $m^{\ast }$. The replacement does not influence the BCS'major
equilibrium predictions. It partly explains why BCS theory that
ignores the effect of lattice periodic potential can still explain
successfully superconductivity. Comparing with bare electron mass,
$m^{\ast }$ has s special property. It may be negative. For
example, the electrons in top of band have the negative $m^{\ast
}$, which is a very important conclusion of band theory. GL theory
can be derived from the microscopic BCS theory. Therefore its mass
may also be negative. The negative $m^{\ast }$ only holds true
near the top of the band, which means that $|-i\hbar \nabla \psi
|^{2}$ should be limited. We need not be afraid that the minimum
of the GL free energy will be unbound from below.

Although the hole description is always introduced when $%
m^{\ast }$ is negative, its application to superconductor is
questioned since the hole is the collective behavior of the whole
band, whereas electron pairing is only involved in a thin layer of
electrons nearby Fermi surface.

Now we can affirm that the lattice periodic potential is very
important to HTS. The anomalous properties about HTS are primarily
due to this effect. Three striking anomalous properties: the
stripe phase, the pseudogap and the energy gap nodes, can been
deduce from GL equation if the lattice periodic potential is
considered. BCS theory and Fermi liquid theory are still fit for
HTS. Some opinions should change. The genuine superconducting
phase transition temperature for HTS is the PG temperature $T_p$
where the order parameter has appeared. The zero-resistance
temperature $T_C$ is just the temperature at which the local
superconducting phase is translated into the globe superconducting
phase. The superconducting phase transition happens also in
insulator.


\newpage
\end{document}